\documentclass[11pt]{article}
\usepackage{amssymb, latexsym, amsbsy, amsfonts}

\newtheorem{definition}{Definition}[section]
\newtheorem{theorem}{Theorem}[section]
\newtheorem{proposition}[theorem]{Proposition}
\newtheorem{lemma}[theorem]{Lemma}
\newtheorem{corollary}[theorem]{Corollary}

\newcommand{\cov}{\mathop{\mathrm{cov}}\nolimits}

\newcommand{\W}{\mathop{\mathrm{{\cal W}}}\nolimits}
\newcommand{\vp}{\mathop{\mathrm{{\varphi}}}\nolimits}

\newcommand{\A}{\mathop{\mathrm{{\cal A}}}\nolimits}

\newcommand{\J}{\mathop{\mathrm{{\cal J}}}\nolimits}
\newcommand{\X}{\mathop{\mathrm{{\cal X}}}\nolimits}

\newcommand{\p}{\mathop{\mathrm{{\cal P}}}\nolimits}

\newcommand{\Dom}{\mathop{\mathrm{Dom}}\nolimits}

    \newcommand {\betrag} [1] {\ensuremath{ \left\vert  #1  \right\vert } } 
    \newcommand {\norm} [2] [] {\ensuremath{ \left\Vert  #2  \right\Vert_{#1} } } 
    \newcommand {\R} {\ensuremath{\mathbb{R}}}
    
    \newcommand {\N} {\ensuremath{\mathcal{N}}}

    \newcommand {\om} {\omega}

    \newcommand {\skalprod} [3] [] {\ensuremath{ \left\langle #2,#3 \right\rangle_{#1}}}

    \newcommand {\Z} {\mathbb{Z}}

    \newcommand {\muT} {\mu_{\scriptscriptstyle T}}

    \newcommand {\mr} {\mathrm}
    \newcommand {\G}  {\ensuremath {\mathcal{G}}}
    \newcommand {\g}  {\ensuremath {\mathsf{G}}}

    \newcommand {\EE} {\mathbb E}

    \renewcommand {\P} {\mathcal{P}}

   \newcommand{\cK} {\mathcal K}
   \newcommand{\cT} {\mathcal T}
   \renewcommand{\Z}{\mathcal{Z}}

\newcommand {\Ga} {\ensuremath{\Gamma}}
\newcommand {\ga} {\ensuremath{\gamma}}
\newcommand {\rh} {\ensuremath{\varrho}}

\makeatletter
\@addtoreset{equation}{section}
\makeatother

\begin{document} 

\title{\Large \bf Gibbs measures on Brownian paths: Theory and applications }
\author{ \small Volker Betz{\thanks{Institute for Biomathematics and Biometry, GSF
Forschungszentrum,
Postfach 1129,  D-85758 Neuherberg, Germany}}, J\'ozsef L\H{o}rinczi and Herbert Spohn \\[0.1cm]
{\it \small Zentrum Mathematik, Technische Universit\"at M\"unchen,} \\
{\it \small Boltzmannstr. 3, 85747 Garching bei M\"unchen, Germany} \\ 
}
\date{}
\maketitle
\begin{abstract} 
\noindent
We review our investigations on Gibbs measures relative to Brownian
motion, in particular the existence of such measures and their path
properties, uniqueness, resp. non-uniqueness. For the case when the 
energy only depends on increments, we present a functional central 
limit theorem. We also explain connections with other work and state 
open problems of interest. 
\end{abstract}

\section{Introduction} \label{Sect1}

The probability measures studied in Statistical Mechanics have the
generic structure
\begin{equation}\label{1.1}
\frac{1}{Z} \exp[-\mathcal{E}]\quad \times \quad \textrm{\textit{a
priori} measure}
\end{equation}
The \textit{a priori} measure is explicit and simple. The energy
function $\mathcal{E}$ is defined on the same space as the
\textit{a priori} measure and the partition function $Z$ makes
(\ref{1.1}) a probability measure. Of course, it is understood
that $\mathcal{E}$ has a natural structure, as dictated by
concrete applications.

One much studied class of examples is that of lattice spin systems 
with finite state space $S$. Then the {\it a priori} measure is the 
product over the lattice points of the counting measure on $S$. The
energy function typically has the form
\begin{equation}
{\cal E}_\Lambda = (k_{\tiny\textrm{B}} T)^{-1}
\sum_{x,y \in \Lambda} U(\sigma_x,\sigma_y, |x-y|)
\label{1.2}
\end{equation}
where $U: S^2 \times \R^+ \to \R$ is a pair potential, and $\sigma_x
\in S$ is the value of the spin at site $x$ of the finite subset
$\Lambda$ of the lattice. The inverse temperature 
$(k_{\tiny\textrm{B}} T)^{-1}$ appears as a strength factor
multiplying the energy.

The specific expression of the measures as formally given by
(\ref{1.1}) is actually firmly grounded in the experience of
rigorous statistical mechanics. At least in the context of lattice spin 
systems with compact state space the so emerging {\it Gibbs measures} 
prove to provide a proper mathematical description of thermodynamic 
equilibrium states and thus they play a fundamental role in the 
theory of phase transitions. In more specific cases, such as the Potts
model, these measures make a strong link between locality properties 
and memory effects (Markov random fields), variational principles 
involving the minimization of free energy so that states appear as 
tangent functionals (large deviation theory), and the understanding 
in terms of percolation properties of how macroscopic long range order 
builds up from small scale events governed by chance (stochastic 
geometry). Although as soon as we leave the class of discrete models 
these relationships are not as clear any longer, these signposts 
pinpoint a programme of a general theory of Gibbs measures from which 
one can take an inspiration. In this paper we present the first steps 
in developing a theory of Gibbs measures on path space. 
  
We will study the case where the \textit{a priori} measure is Brownian 
motion in $\mathbb{R}^d$. Let us denote by $t \to X_t \in
\mathbb{R}^d$ a Brownian path and by $\mathcal{W}$ the Wiener
measure. Since $t \in \mathbb{R}$, in the parlance of Statistical 
Mechanics our model is one-dimensional with $d$ components. The finite 
box $\Lambda$ corresponds to the time interval $[-T,T]$. $\mathcal{W}$ 
has then to be supplied with appropriate boundary conditions. For 
example one could pin the path at both endpoints, $X_{-T}=0$, $X_T=0$, 
in which case $\mathcal{W}$ would turn into a Brownian bridge. The 
simplest energy function is given through an ``on site'' potential $V:
\mathbb{R}^d \to \mathbb{R}$ and takes the form
\begin{equation}\label{1.3}
\mathcal{E}_{1,T} = \int^T_{-T} V(X_t)dt.
\end{equation}
The analogue of the pair interaction energy (\ref{1.2})
transcribes as
\begin{equation}\label{1.4}
\mathcal{E}_{2,T} = \int^T_{-T} \int^T_{-T}W(X_t,X_s,t-s)dtds
\end{equation}
with $W:\mathbb{R}^{2d} \times \mathbb{R} \to \mathbb{R}$,
$W(x,x',t)=W(x',x,t)$, and $\int |W(x,x',t)| dt < \infty$. In
Statistical Mechanics energies are proportional to the volume,
i.e. proportional to $T$ in our case. Clearly, in spirit this is
satisfied by both energies (\ref{1.3}) and (\ref{1.4}). With these
preparations a Gibbs measure on path space reads as
\begin{equation}\label{1.5}
\frac{1}{Z(T)} \exp\big[-\mathcal{E}_{1,T} (X) - \mathcal{E}_{2,T}
(X)\big] \delta(X_{-T})\delta(X_T) d\mathcal{W}(X)\,.
\end{equation}
Of course, there is considerable freedom in how to pick the energy
function. (\ref{1.3}) and (\ref{1.4}) come up naturally from
applications. A further set of examples is obtained by replacing
in (\ref{1.3}), (\ref{1.4}) the Riemann integrals by stochastic
integrals as
\begin{equation}\label{1.6}
\mathcal{\widetilde{E}}_{1,T}(X)= \int^T_{-T}a(X_t) \cdot
dX_t\,,\quad \mathcal{\widetilde{E}}_{2,T}(X)= \int^T_{-T}
\int^T_{-T}dX_s \cdot W(X_t,X_s,t-s)dX_t\,,
\end{equation}
with $a(x)$ a vector field and $W(x,x',t)$ a $d \times d$ matrix.
Since our own work is centered more around (\ref{1.5}), we will
concentrate exclusively on this case.

Our plan is first to explore the probabilistic structure. In the
final chapter we list various applications for which measures of
the form (\ref{1.5}) with specific choices of $\mathcal{E}_1$ and
$\mathcal{E}_2$ appear. From there it will also be apparent that
each application poses specific questions not covered by general
theory.

Broadly speaking, given the measure in (\ref{1.5}) there are two
limiting procedures of interest.

\vspace{0.2cm}
\noindent
\textit{i) Short distance (ultraviolet) limit.} The box $[-T,T]$
is fixed and the interaction is singular on the diagonal. The
prototype are polymer measures, where self-crossings are penalized 
by the energy
\begin{equation}\label{1.7}
\mathcal{E}_{T,\tiny\textrm{poly}} (X) = \int^T_{-T} \int^T_{-T}
\delta_n (X_t-X_s) dtds\,.
\end{equation}
Here $\delta_n \geq 0$ with support in a ball of radius $1/n$
centered at the origin. One goal is then to prove that the Gibbs
measure in (\ref{1.5}) with the energy (\ref{1.7}) has a limit as
$n \to \infty$. Problems of these type also come up in proving
renormalizability of quantum field theories. They have been
studied in considerable detail. We refer to \cite{gall,ros1,ros2,
var,bolt} and references therein. A more detailed discussion is 
outside the scope of the present review and we will always assume 
that $W$ is locally bounded.

\vspace{0.2cm}
\noindent
\textit{ii) Large distance (infinite volume) limit.} The goal is
to show that the measure in (\ref{1.5}) has a limit as $T \to
\infty$. The limit measure has then conditional expectations \`{a}
la Dobrushin, Lanford, and Ruelle. As standard in the theory of
Gibbs measures, the issue divides into the existence of a limit
measure and the dependence of the limit measure on the choice of
boundary conditions.

\vspace{0.2cm}
The infinite volume limit will be discussed in Section \ref{Sect2}. 
A prerequisite is the case $W \equiv 0$, which leads
to the theory of $P(\phi)_1$-processes, i.e. reversible diffusion
processes with constant diffusion, which will be taken up in
Section \ref{Sect2.1}. If the interaction $W$ is weak, one expects 
that the qualitive properties of the stationary $P(\phi)_1$-process 
remain intact. Technically, a cluster expansion will be used to 
establish such a result. The basic set-up will be explained in Section 
\ref{Sect2.2}. It differs from the more convential cluster expansions 
because the \textit{a priori} measure is not a product measure and the
configurations are segments of Brownian paths rather than the better
understood $\mathbb{R}$- or $\mathbb{Z}$-valued spins. To prove 
existence of the limit measure with no restriction on the
interaction strength requires other methods. One possibility is 
domination and monotonicity \cite{os}. In Section \ref{Sect2.3} we 
explain a more general scheme, which relies on having an
essentially bounded interaction energy between the path $\{X_t, t
\leq 0\}$ and the path $\{X_t, t \geq 0\}$. Under such a condition
we prove that the Gibbs measure is unique, i.e. independent of the
choice of boundary conditions within a reasonable class. To have
non-uniqueness, the interaction energy must increase at least as
$\log T$, or equivalently $W(x,x',t)$ has to decay at least as
slow as $|t|^{-2}$ for large $|t|$. In Section \ref{Sect2.4} we 
discuss a specific example, for which it can be shown that the limit 
measure depends on the choice of the boundary conditions.

Another case of interest is the energy
\begin{equation}\label{1.8}
\mathcal{E}(X)= \int^T_{-T} \int^T_{-T} W(X_t-X_s, t-s) dtds
\end{equation}
with $\int |t|W(x,t)dt < \infty$, hence zero external potential
$V$. As the expression shows, the energy depends only on path 
increments. Thus one expects that, under the Gibbs measure 
(\ref{1.5}) for ${\cal E}_{1,T} + {\cal E}_{2,T}$ replaced by
${\cal E}$, $X_t$ behaves like Brownian motion with some effective 
diffusion coefficient. For example, if $X$ is pinned as $X_{-T} = 0 
= X_T$, then $\mathbb{E}_T (X^2_0)\simeq T$, at large $T$. The $T 
\to \infty$ limit of the measure (\ref{1.5}) will not exist and the 
more sensible project is to prove an invariance principle under
suitable rescaling. This will be explained in Section \ref{Sect3}. 
Finally, in Section \ref{Sect4} we discuss some specific
applications. 

At this point we would like to take the opportunity to thank the
organizers of the SPP 1033 ``Interacting Stochastic Systems of
High Complexity" for their initiative. The Schwerpunkt turned out
to be a successful enterprise for joint research in the applied
areas of probability theory.

\section{Gibbs measures}
\label{Sect2}
\subsection{The case of external potential}
\label{Sect2.1}
First we outline a method on how to represent $P(\phi)_1$-processes
(i.e., Brownian motion in the presence of an external potential) in 
terms of Gibbs measures. Since $B_t$, the outcomes of Brownian motion,
are correlated for different values of $t$, Wiener measure carries
some dependence and is not as simple as a product measure. However, 
by its Markovianness and since this property survives under the
potentials we consider, $P(\phi)_1$-processes are tractable to a fair
extent, which is a first step toward understanding more complicated
cases, such as (\ref{1.4}) when also a pair interaction in present. 
For early results we refer to \cite{Sim1,Sim2}, for details of 
Gibbsian description as well as proofs and a discussion of the related
literature see \cite{BL}; the arguments used here are largely based on 
a spectral theoretic analysis. 

Denote $V^+ = \sup\{0,V\}$ and $V^- = \inf\{-V,0\}$. Two classes of
external potential $V: \R^d \to \R$ will be considered: 
\begin{trivlist}
\item
\hspace{0.2cm} (V1) {\it Kato-class.} \hspace{0.2cm}
Here $V^- \in K_d$ and $V^+ \in K_d^{\mathrm{\tiny{loc}}}$, with
\begin{eqnarray*}
&& K_1 = \{ V: \sup_{x \in \R} \int_{\betrag{x-y} \leq 1}
\betrag{V(y)} \, dy < \infty \}, \\
&& K_d = \{ V: \lim_{r \to 0} \sup_{x \in \R^d} \int_{\betrag{x-y}
  \leq r} \betrag{V(y)} q(\betrag{x-y}) \, dy = 0\} \;\;\; \mbox{if} 
  \;\;\; d \geq 3,
\end{eqnarray*}
with $q(x) = -\log |x|$ for $d=2$, and $q(x) = 1/|x|^{d-2}$ for $d
\geq 3$, and the local Kato-class 
\begin{equation}
K_d^{\mathrm{loc}} = \left\{ f: f 1_A \in K_d \;\;\; \mbox{for each
    compact} \;\;\; A \subset \R^d \right\}.
\end{equation}
\item
\hspace{0.2cm} (V2) {\it Confining potentials.} \hspace{0.2cm}
$V$ is bounded from below and continuous, moreover $V(x) = 
a|x|^{2s} + o(|x|^{2s})$, with some $s > 1$ and $a > 0$.
\end{trivlist}
Examples of Kato-class potentials include smooth functions bounded
from below, but also some local (e.g. Coulomb) singularities are
allowed. In particular, (V2) is a specific case of (V1). The sets 
$K_d$ can also be characterized in terms of Wiener integrals. 

For $V$ having either of the regularity properties above define the
Schr\"odinger operator $H = -1/2 \Delta + V(x)$ on $L^2(\R^d,dx)$ as 
a sum of quadratic forms ($V$ is regarded as a multiplication
operator). Then $C^\infty_0(\R^d)$ is a form core on which $H$ is 
essentially self-adjoint and bounded from below. If the bottom of 
the spectrum $E_0$ of $H$ is a simple eigenvalue, then the
corresponding eigenfunction $\psi_0$ (ground state) is strictly 
positive. The semigroup $e^{-tH}$, $t \geq 0$, exists on 
$L^2(\R^d, dx)$, and it is an integral operator with positive, 
continuous, uniformly bounded kernel $G_t(x,y)$. For (V2)-type 
potentials the semigroup is moreover intrinsically ultracontractive. 
That is, with the probability measure $d\nu = \psi_0^2 dx$ on $\R^d$, 
and isometry $j: L^2(\R^d,d\nu) \rightarrow L^2(\R^d,dx)$, $f \mapsto 
\psi_0 f$, the operator 
\begin{eqnarray}
H_\nu f &=& (j^{-1} (H-E_0) j) f = \frac{1}{\psi_0} (H-E_0)(\psi_0 f) 
\nonumber \\
&=& -\frac{1}{2} \Delta f - \left({\nabla \ln \psi_0},\nabla f \right)
_{\R^d}, 
\end{eqnarray}
with $\Dom H_\nu = j^{-1}(\Dom H)$, defines a semigroup $e^{-tH_\nu}$ 
for all $f \in L^2(\R^d,d\nu)$ and $t \geq 0$. Intrinsic 
ultracontractivity of $e^{-tH}$ means that $e^{-tH_\nu}$ is 
ultracontractive, i.e. it maps $L^2(\R^d, d\nu)$ into 
$L^\infty(\R^d,d\nu)$ continuously, or equivalently, 
$||e^{-tH_\nu}||_{2,\infty} < \infty$, $\forall t \geq 0$. 

\vspace{0.2cm}
Choose now $H$ to be a Schr\"odinger operator such that its ground 
state $\psi_0$ exists. For convenience and without loss we shift the 
potential by $E_0$ so that the bottom of the spectrum of $H$ is $0$.
For $t_1 < \ldots < t_n \in \R$, $f_1, \ldots, f_n \in L^2(\R^d,dx) 
\cap L^{\infty}(\R^d,dx)$, the {\it $P(\phi)_1$-process} associated 
with $H$ is the unique probability measure $P$ on path space 
$C(\R,\R^d)$ defined by
\begin{eqnarray} 
\lefteqn{\int f_1(X_{t_1}) \ldots f_n(X_{t_n}) dP(X) } \nonumber \\
& = & (\psi_0 f_1, 
e^{-(t_2-t_1)H}f_2 \ldots e^{-(t_n - t_{n-1})H}
 f_n \psi_0)_{L^2(\R^d,dx)}.
\label{pfi}
\end{eqnarray}
$P$ is indeed a probability measure as $e^{-tH}\psi_0 = \psi_0$ and 
$\norm[2] {\psi_0} = 1$. A $P(\phi)_1$-process is a reversible 
stationary Markov process with stationary measure $d\nu$ and generator 
$H_{\nu}$, and it has almost surely continuous paths. It is moreover 
the stationary solution of the stochastic differential equation 
(It\^o-diffusion) $$ dX_t = (\nabla \ln {\psi_0})(X_t) \, dt + dB_t,$$
where $B_t$ denotes Brownian motion on $\R^d$.

\vspace{0.2cm}
Processes of this type can be given a Gibbsian description. We
emphasize that since in the present stage there are no useful
relationships available with variational principles etc as discussed
in the Introduction, here the basic fact is that there is at all a 
probability measure associated with the scalar product in (\ref{pfi}),   
a consequence of the Riesz-representation theorem, while its 
Gibbsianness comes second to it. That we are able to identify this
measure as a Gibbs measure leads however to further insight.
 
Denote $\X 
= C(\R,\R^d)$, the space of continuous functions from $\R$ to $\R^d$, 
and its $\sigma$-field $\A = \sigma (\pi_t: t \in \R)$ generated by
the point evaluations $\pi_t: \X \to \R^d$, $X \mapsto \pi_t(X) =
X_t$. These will be the configuration space and $\sigma$-field for the 
Gibbs measure, respectively. For $[-T,T] \subset \R$ we denote by $\A_T$
the $\sigma$-field $\sigma (\pi_t: t \in [-T,T]) \subset \A$; also, we 
put $[-T,T]^c = \R \smallsetminus [-T,T]$. 

Write as before $\W$ for Wiener measure, and $\W_{[-T,T]}^{\xi,\eta}$ 
for the Wiener measure conditional on starting in $\xi$ at time $-T$ 
and ending in $\eta$ at time $T$. This Brownian bridge can be extended 
to a measure on $\X$ by picking an $Y \in \X$ and putting $\W_T^{Y} = 
\W_{[-T,T]}^{Y_{-T},Y_T} \otimes \; \delta_{[-T,T]^c}^Y$, with Dirac 
measure on $C([-T,T]^c,\R^d)$ concentrated on $Y |_{[-T,T]^c}$. 
$\W_T^Y$ is thus a finite measure on $(\X,\A)$; it will serve as 
reference measure for the Gibbs measure to be constructed. 
 
Take any $A \in \A$ and consider
\begin{equation} \label{(3.0)}
dP_T(A|Y) = \frac{1}{Z_T(Y)} 1_A(X) e^{- \int_{-T}^T V(X_s) ds} \, 
d\W^Y_T(X),
\end{equation}
where
\begin{equation} \label{(3.0.5)}
Z_T(Y) = \int e^{- \int_{-T}^T V(X_s) ds} \, d\W^Y_T(X)
\end{equation}
is the partition function turning $P_T$ into a probability measure. 
\begin{definition} \label{3.2}
Let $\X^* \subset \X$. A probability measure $\cal P$ on $(\X,\A)$ is
called a {\rm Gibbs measure} for potential $V$ and reference measure 
$\W^Y$, if for every bounded interval $[-T,T] \subset \R$ 
\begin{enumerate}
\item
${\cal P}|_{{\cal A}_T} \; \ll \; \W|_{{\cal A}_T}$, 
\item
for every $A \in \A$ the function $Y \mapsto {\cal P}_{T}(A|Y)$ given 
by the right hand side of (\ref{(3.0)}) is a regular version of the 
conditional probability ${\cal P}(A|\A_{[-T,T]^c})$. 
\end{enumerate}
A probability measure ${\cal P}_T$ on $([-T,T],\A_T)$ is called a 
{\rm finite time interval Gibbs measure} for $V$ and reference
measure $W^Y_T$ if for every bounded interval $[-S,S] \subset [-T,T]$ 
the function $Y \mapsto {\cal P}_S (A|Y)$ as above is a regular 
version of the conditional probability ${\cal P}_S(A|\A_{[-S,S]^c})$. 
Furthermore, a Gibbs measure $\cal P$ is said to be supported by
$\X^*$ whenever ${\cal P} (\X^*)=1$.
\end{definition}
This definition rests on the DLR conception of Gibbs measure. In this 
sense we then have 
\begin{theorem} \label{3.1}
A $P(\phi)_1$-measure $P$ corresponding to potential $V$ is a Gibbs 
measure with respect to $V$ and Wiener measure.
\end{theorem}
On the other hand, that $P_T(\cdot|Y)$ are a family of finite time 
interval Gibbs measures indexed by bounded intervals can be seen in a 
straightforward way. It can be proven by a monotone class argument
that (infinite time interval) Gibbs measures on path space can be 
obtained by limits of finite interval Gibbs measures, similarly to 
the case known from lattice spin models. In this limiting procedure
thus one must have a control of boundary conditions.

\vspace{0.2cm}
A Gibbs measure associated with a $P(\phi)_1$-process need not be
unique. This non-uniqueness appears as a dependence of the Gibbs
measure on the boundary conditions. An example showing this is the 
Ornstein-Uhlenbeck process, in which case uncountably many Gibbs 
measures can occur for the same potential. This is related with the 
rate how the boundary paths increase, or in other words, how fast 
for each $T$ the boundary path on $[-T,T]^c$ has to ``forget'' 
that it was free Brownian motion before stepping in $[-T,T]$ 
where it must ``steady down'' to conform with the correct distribution 
prescribed by (\ref{(3.0)}). 
\if
Take $V_{\mathrm{\tiny{OU}}}(x) = \frac{1}{2} (x^2-1)$, $x \in \R$,
and consider the corresponding Schr\"odinger operator 
$H_{\mathrm{\tiny{OU}}}$. As well-known, its ground state is
$\psi_0(x) = \pi^{-1/4} e^{-x^2/2}$, and the $P(\phi)_1$-process 
corresponding to it is the one-dimensional Ornstein-Uhlenbeck 
process. Moreover, Mehler's formula gives explicitly the integral 
kernel of $e^{-tH_{\mathrm{\tiny{OU}}}}$. Fix $\alpha, \beta \in \R$ 
and define for $s,x \in \R$
\begin{eqnarray*}
&& \psi^l_s(x) = \pi^{-1/4} \exp \left( -\frac{1}{2}(x + 
\alpha e^{-s})^2 \right) 
\exp \left( \frac{\alpha e^{-s}}{2} \right)^2,\\
&& \psi^r_s(x) =  \pi^{-1/4} \exp \left( -\frac{1}{2}(x + 
\beta e^{s})^2 \right) 
  \exp \left( \frac{\beta e^{s}}{2} \right)^2.
\end{eqnarray*}
A simple calculation shows that $e^{-tH_{\mathrm{\tiny{OU}}}} 
\psi_s^l = \psi_{s+t}^l$, $e^{-tH_{\mathrm{\tiny{OU}}}}\psi_s^r = 
\psi_{s-t}^r$, and $(\psi_s^l,\psi_s^r) = e^{\alpha \beta / 2}$. 
Thus $\int f_1(X_{t_1})...f_n(X_{t_n}) dP_{\alpha,\beta}(X) = 
e^{-\alpha\beta/2} (\psi_{t_1}^l f_1, e^{-(t_2-t_1)
H_{\mathrm{\tiny{OU}}}}f_2 \ldots e^{-(t_n-t_{n-1})
H_{\mathrm{\tiny{OU}}}} f_n \psi_{t_n}^r)$ defines the finite 
dimensional distributions of a probability measure $P_{\alpha,\beta}$ 
on $C(\R,\R)$ of a Gaussian Markov process, which is stationary if 
and only if $\alpha = \beta = 0$. Furthermore, it is easily seen that 
$P_{\alpha,\beta}$ is a Gibbs measure for every $\alpha, \beta \in
\R$. In this case thus uncountably many Gibbs measures exist for the 
same potential. 
\fi
A condition for uniqueness of the Gibbs measure is provided by the 
following theorem. 
\begin{theorem} \label{4.3}
Let $H$ be a Schr\"odinger operator for a Kato-class potential $V$ 
such that the spectral gap $\Lambda$ of $H$ is strictly positive, and 
let $\psi_0$ be its ground state. Put
\begin{equation} \label{(4.3.0)}
\X^* = \{ X \in \X: \lim_{|t|\to\infty} \frac{e^{-\Lambda 
\betrag{t}}}{\psi_0(X_t)} = 0 \}.
\end{equation}
Then the $P(\phi)_1$-measure $P$ corresponding to $V$ is the unique 
Gibbs measure for $V$ supported by $\X^*$. If, furthermore, $V$ is a
$(V2)$-type confining potential, then $P$ is the unique Gibbs measure 
supported on the entire $\X$. 
\end{theorem}
The first part of the statement results from an argument using direct
estimates, the second relies on ultracontractivity.

By restricting to (V2)-type potentials and making use of the fact that 
for this class $\psi_0$ is bounded both from below and above by $C 
\exp(-\theta |x|^{s+1})$, with suitable constants $C, \theta > 0$ for 
the two bounds respectively, we obtain from Theorem \ref{4.3} that 
those paths are typical for the $P(\phi)_1$-measure that grow 
asymptotically like $t^{1/(s+1)}$.

\subsection{Weak pair potential: cluster expansion}
\label{Sect2.2}

Next we turn to discussing whether Gibbs measures can be defined 
also for Brownian motion subjected to both an external and a pair 
interaction potential. Such a process is not Markovian and therefore 
not accessible to spectral analysis. Instead, we will develop a 
cluster expansion; for details and proofs see \cite{LM}.

We use the same set-up as before. The pair interaction potential is a
measurable function $W: \R^d \times \R^d \times \R \to \R$ with the 
(inessential) symmetry properties $W(\cdot,\cdot,t-s) = 
W(\cdot,\cdot,|t-s|)$, $W(x,y,\cdot) = W(y,x,\cdot)$, $x,y \in \R^d$, 
$s,t \in \R$, and satisfying either of the following regularity 
conditions:
\begin{trivlist}
\item
\hspace{0.2cm} (W1) There is $R > 0$ and $\alpha > 2$ such that 
\begin{equation}
|W(x,y, t-s)| \;\leq\; R \; \frac{|x|^2 + |y|^2}{1 + |t-s|^\alpha}
\label{a21}
\end{equation}
for every $x,y \in \R^d$ and $t,s \in \R$.
\item
\hspace{0.2cm} (W2) There is $R > 0$ and $\alpha > 1$ such that 
\begin{equation}
|W(x,y, t-s)| \;\leq\; \frac{R}{1 + |t-s|^\alpha}
\label{a22}
\end{equation}
for every $s,t \in \R$ and $x,y \in \R^d$.
\end{trivlist}

\vspace{0.2cm}
\noindent
For $[-T,T] \subset \R$ write
\begin{equation}
W_{[-T,T]}(X|Y) = W_{[-T,T]}(X) + W_{[-T,T]}^Y(X) 
\end{equation}
for the ``total energy'' associated with configuration $X \in
\X_{[-T,T]}$ given the boundary configuration $Y = Y^- \cup 
Y^+$, with $Y^- \in \X_{(-\infty,-T]}$ resp. $Y^+ \in
\X_{[T,\infty)}$. Term by term,
\begin{equation}
W_{[-T,T]}(X) = \int_{-T}^{T}\int_{-T}^{T} 
W (X_t, X_s, s-t) ds dt 
\label{inn}
\end{equation}
is the ``internal energy'' associated with the path inside 
$[-T,T]$, and
\begin{eqnarray}
W_{[-T,T]}^Y(X) & = & 2 \int_{-\infty}^{-T} dt \int_{-T}^{T} 
ds \; W(Y^-_t, X_s,t-s) + \nonumber \\ 
&& +  2 \int_{T} ^\infty dt \int_{-T}^{T} 
ds \; W(Y^+_t,X_s,t-s)
\label{int}
\end{eqnarray}
is the ``interaction energy'' between $X$ and the boundary 
path $Y$. We calibrate the interaction energy such that 
$W_{[-T,T]}^0(X) = 0$. As before, $P_T(\;\cdot\;|Y) = 
P_{[-T,T]}(\;\cdot\;|Y^-_{-T},Y^+_{T})$ is the conditional 
distribution of the reference measure for the given boundary 
condition $Y$ (which by Markovianness obviously depends only 
on the positions attained at the ends of the interval). It is 
readily checked that $\mu_{[-T,T]}(\;\cdot\;|Y)$, with $Y \in 
C([-T,T]^c, \R^d)$, is a family of finite time interval Gibbs 
measures. We also allow $\lambda \in \R$, a parameter which 
can be interpreted as the strength of the coupling of the pair 
interaction to the Brownian paths. 

Consider now a $P(\phi)_1$-process with stationary measure 
$d\nu = \psi_0^2 dx$ and transition probability density 
\begin{equation}
g_t(x|y) = \frac{\psi_0(x) G_t(x,y)}{\psi_0(y) e^{-E_0 t}}.
\label{tp}
\end{equation}
Denote again the probability distribution of this process by $P$, 
and by $P_T$ its restriction to the field ${\cal A}_{[-T,T]}$. We 
take this as reference measure in constructing the finite time 
interval Gibbs measures on $\X_{[-T,T]}$ for the pair potentials 
above: 
\begin{equation}
d\mu_T(A|Y) = \frac{1}{\Z_T(Y)} 1_A(X) 
e^{-\lambda W_{[-T,T]}(X|Y)} dP_T(X|Y),
\label{wgibb}
\end{equation}
for any $A \in \A$ and boundary condition $Y$. Here we speak 
about Gibbs measure $\mu$ in the same sense as in Definition 
\ref{3.2}, now for potential $W$ and reference measure $P$. 
The partition function is 
\begin{equation}
\Z_T(Y) = \int e^{-\lambda W_{[-T,T]}(X|Y)} dP_T(X|Y). 
\label{wz}
\end{equation}
As said before, Gibbs measures can be obtained as limits over finite
time interval Gibbs measures. Thus it is of interest whether the 
sequence of Gibbs measures $\mu_T$ has a limit as $T \to \infty$; if 
it does then it provides a Gibbs measure on the full path space as 
soon as also condition (i) of Definition \ref{3.2} is met.

\begin{theorem}
Suppose $V$ and $W$ satisfy assumptions (V2), respectively either (W1)
or (W2). Take any unbounded increasing sequence $T^{(n)}$ of positive 
real numbers, and suppose $0 < |\lambda| \leq \lambda^*$ with 
$\lambda^*$ small enough. Then the local weak limit $\lim_{n
  \rightarrow \infty} \mu_{T^{(n)}} = \mu$ exists and is a Gibbs 
probability measure on $(\X,{\cal A})$ with respect to $W$ and
reference measure $P$. Moreover, $\mu$ does not depend on the choice
of sequence $T^{(n)}$. 
\end{theorem}

\vspace{0.2cm}
In order to prove this convergence we use a cluster expansion
controlled by the small parameter $\lambda$. Next we sketch the 
cluster representation of the partition function (\ref{wz}) and
outline the main steps of the proof. For simplicity we start with 
free boundary conditions, i.e. $Y = 0$ in (\ref{wgibb}). 

Take a division of $[-T,T]$ into disjoint intervals $\tau_k = (t_k, 
t_{k+1})$,  $k = 0,..., N-1$, with $t_0 = -T$ and $t_N = T$, each of 
length $b$, i.e. fix $b = 2T/N$; for convenience we choose $N$ to be 
an even number so that the origin is endpoint to some intervals. We 
break up a path $X$ into pieces $X_{\tau_k}$ by restricting it to 
$\tau_k$. The total energy contribution of the pair interaction then
becomes
\begin{equation}
W_T := \int_{-T}^T\int_{-T}^T W (X_t, X_s, s-t) ds dt = 
\sum_{0 \leq i < j \leq N-1} W_{\tau_i,\tau_j}
\label{sum}
\end{equation}
where with the notation $\J_{ij} = \int_{\tau_i} dt \int_{\tau_j} 
W(X_s, X_t, s-t) ds$ we have
\begin{equation}
W_{\tau_i,\tau_j} 
= \left\{ 
\begin{array}{ll}
\vspace{0.2cm} 
\J_{ij} + \J_{ji} & \mbox{if $|i-j| \geq 2$} \nonumber \\
\vspace{0.2cm} 
\frac{1}{2} (\J_{ii} + \J_{jj}) + \J_{ij} + \J_{ji}  
& \mbox{if $|i-j| = 1$, and 
$i \neq 0$, $j \neq N - 1$} \nonumber \\
\vspace{0.2cm} 
\J_{ij} + \J_{ji} + \frac{1}{2} \J_{00} 
& \mbox{if $i = 0$ and $j = 1$} \nonumber \\ 
\vspace{0.2cm} 
\J_{ij} + \J_{ji} + \frac{1}{2} \J_{N-1 \; N-1} 
& \mbox{if $i = N-1$ and $j = N-2$} \nonumber \\
\end{array} \right.
\end{equation}
(For keeping the notation simple we do not make explicit the 
$X$ dependence of these objects.) 
By using (\ref{sum}) we obtain
\begin{equation}
e^{-\lambda W_T} = \prod_{0 \leq i < j \leq N-1} 
(e^{-\lambda W_{\tau_i,\tau_j}} + 1 - 1) =  1 + \sum_{{\cal R} \neq 
\emptyset} \prod_{(\tau_i,\tau_j) \in {\cal R}} 
(e^{-\lambda W_{\tau_i,\tau_j}} - 1).
\label{sum1}
\end{equation}
Here the summation is performed over all nonempty sets of different 
pairs of intervals, i.e. ${\cal R} = \{(\tau_i,\tau_j): 
(\tau_i, \tau_j) \neq (\tau_{i'},\tau_{j'}) \; \mbox{whenever} \; (i,j) 
\neq (i',j') \}$. 

In order to keep this and the forthcoming summations in hand we need 
a few more notations. Two distinct pairs of intervals
$(\tau_i,\tau_j)$ and $(\tau_{i'},\tau_{j'})$ will be called 
{\it directly connected} and denoted $(\tau_i,\tau_j) \sim 
(\tau_{i'},\tau_{j'})$ if one interval of the pair $(\tau_i,\tau_j)$ 
coincides with one interval of the pair $(\tau_{i'},\tau_{j'})$. A 
set of connected pairs of intervals is a collection 
$\{(\tau_{i_1},\tau_{j_1}),..., (\tau_{i_n},\tau_{j_n})\}$ in which 
each pair of intervals is connected to another through a sequence of 
directly connected pairs, i.e., for any $(\tau_i,\tau_j) \neq 
(\tau_{i'},\tau_{j'})$ there exists $\{(\tau_{k_1},\tau_{l_1}),..., 
(\tau_{k_m},\tau_{l_m})\}$ such that $(\tau_i,\tau_j) \sim 
(\tau_{k_1},\tau_{l_1}) \sim ... \sim (\tau_{k_m},\tau_{l_m}) 
\sim (\tau_{i'},\tau_{j'})$. A maximal set of connected pairs of 
intervals is called a {\it contour} and denoted by $\gamma$. We denote 
by $\bar\gamma$ the set of all intervals that are elements of the
pairs of intervals belonging to contour $\gamma$, and by $\ga^*$ the 
set of time-points of intervals appearing in $\bar\ga$. Clearly, $\cal 
R$ can be decomposed into maximal connected components, i.e. contours: 
${\cal R} = \{\ga_1,...,\ga_r\}$ with $\bar\ga_i \cap \bar \ga_j = 
\emptyset$, $i \neq j$; $i,j = 1,...,r$. 

The sum in (\ref{sum1}) is then further expanded as
\begin{equation}
\sum_{{\cal R} \neq \emptyset} \prod_{(\tau_i,\tau_j) \in \cal R} 
(e^{-\lambda W_{\tau_i,\tau_j}} - 1) = \sum_{r \geq 1} 
\sum_{\{\gamma_1,...,\gamma_r\}} \prod_{k = 1}^r 
\prod_{(\tau_i,\tau_j) \in \gamma_k} (e^{-\lambda W_{\tau_i,\tau_j}} 
- 1)
\label{sum2}
\end{equation}
where now summation goes over collections $\{\gamma_1,...,\gamma_r\}$ 
of contours such that $\bar\ga_k \cap \bar\ga_{k'} = \emptyset$ 
unless $k = k'$. 
                                
A collection of consecutive intervals $\{\tau_{j},\tau_{j+1}...,
\tau_{j+k} \}$, $j \geq 0$, $j+k \leq N-1$ is called a {\it chain}. 
As in the case of contours, $\bar \rh$ and $\rh^*$ mean the set of 
intervals belonging to the chain $\rh$ and the set of time-points in 
$\rh$, respectively. We call two contours $\ga_1, \ga_2$ disjoint if 
they have no intervals in common, i.e. $\bar \ga_1 \cap \bar \ga_2 = 
\emptyset$. Two chains $\rh_1, \rh_2$ are called disjoint if they have 
no common time-points, i.e. $\rh_1^* \cap \rh^*_2 = \emptyset$. Take
now a non-ordered set of disjoint contours and disjoint chains, 
$\Ga = \{\gamma_1,...,\gamma_r;\varrho_1,...,\varrho_s\}$, with some 
$r \geq 1$ and $s \geq 0$. Note that such contours and chains may have 
common time-points. We use the notation $\Ga^* = (\cup_i \ga^*_i) \cup 
(\cup_j \rh_j^*)$ for the set of all time-points appearing as
beginnings or ends of intervals belonging to some contour or chain in 
$\Gamma$. Also, we put $\bar \Ga = (\cup_i \bar\ga_i) \cup (\cup_j 
\bar\rh_j)$ for the set of intervals appearing in $\Ga$ through 
entering some contours or chains. Denote by $\partial^-\rh$ resp. 
$\partial^+\rh$ the leftmost resp. rightmost time-points belonging to 
$\rh$. $\Gamma$ is called a {\it cluster} if $\{ \ga^*_1,..., \ga^*_r;
\rh^*_1,...,\rh^*_s\}$ is a connected collection of sets and for every 
$\varrho \in \Gamma$ we have that $\partial^-\rh, \partial^+\rh \in 
\cup_{j=1}^r\ga^*_j$. This means that in a cluster chains have no
loose ends.

Next we fix the positions of path $X$ at the time-points of the 
division, i.e. we put $X_{t_k} = x_k$, for all $k = 0,...,N$, with $-T 
= t_0 < t_1 < ... < t_N = T$. The distribution of path $X$ in interval 
$[-T,T]$ conditional on the positions attained at the fixed times is
\begin{equation}
dP_T(X_{\tau_0},...,X_{\tau_{N-1}} | X_{t_0} = x_0, \ldots , 
X_{t_N} = x_N) = \prod_{k=0}^{N-1} dP_{\tau_k}(X_{\tau_k} | x_k,
x_{k+1}).
\label{mar}
\end{equation}
We use the shorthand at the right hand side for the corresponding 
conditional probabilities for easing the notation. Let
$p_{t_0,...,t_N} (x_0,...,x_N)$ be the density with respect to 
$\prod_{k=0}^N d\nu_k(x_k)$ of the joint distribution of positions 
of path $X$ recorded at the time-points $t_0,...,t_N$. Here $d\nu_k$ 
denotes a copy of $d\nu$ for each $k = 0,...,N$. By Markovianness it 
then follows that
\begin{eqnarray}
p_{t_0,...,t_N} (x_0,...,x_N) 
& = &
\prod_{k=0}^{N-1} g_b(x_{k+1}|x_k) = \prod_{k=0}^{N-1} 
(g_b(x_{k+1}|x_k) - 1 + 1)  
\nonumber \\ 
& = &
1 + \sum_{\cal S} \prod_{k: \tau_k \in {\cal S}} (g_b(x_{k+1}|x_k) - 1). 
\nonumber
\end{eqnarray}
The summation is extended over all nonempty sets ${\cal S} = \{\tau_k
= (t_k,t_{k+1})\}$ of different pairs of consecutive time-points. In a 
similar way as before the latter formula can be recast in the form
\begin{equation}
\sum_{\cal S} \prod_{k: \tau_k \in {\cal S}} \left( g_b(x_{k+1}|x_k) -
  1 \right) = 
\sum_{s \geq 1} \sum_{\{\varrho_1,...,\varrho_s\}} \prod_{j=1}^s 
\prod_{k: \tau_k \in \varrho_j} \left(g_b(x_{k+1}|x_k) - 1 \right).
\label{sum3}
\end{equation}
Here $\{\varrho_1,...,\varrho_s\}$ is a collection of disjoint chains,
and this formula explains the way we defined them before. 

For every cluster $\Ga = \{\ga_1,...,\ga_r;\rh_1,...,\rh_s\}$ define
the function
\begin{equation}
\kappa_\Ga = \prod_{l=1}^r \prod_{(\tau_i,\tau_j) \in \gamma_l} 
(e^{-\lambda W_{\tau_i\tau_j}} - 1) \prod_{m=1}^s 
\prod_{k: \tau_k \in \rh_m}\left(g_b(x_{k+1}|x_k) - 1\right). 
\label{kappa}
\end{equation}
Also, introduce the auxiliary probability measure on $\X_T$
\begin{equation}
d\p_T(X) = \prod_{k=0}^{N-1} dP_{\tau_k}(X_{\tau_k}|x_k,x_{k+1}) 
\prod_{k=0}^N d\nu_k(x_k)
\label{auxi}
\end{equation}
and look at 
\begin{equation}
K_\Ga = \mathbb{E}_{\p_T} [\kappa_\Ga].
\label{aux}
\end{equation}
Note that $\int(g_b(x_{k+1}|x_k)-1) d\nu(x_{k+1}) = 
\int(g_b(x_{k+1}|x_k) -1) d\nu(x_k) = 0$. This is the reason why from
a cluster we rule out chains having loose ends; for any such chain 
$\mathbb{E}_{\p_T} [\kappa_\Ga] = 0$.
 
By putting (\ref{sum2}), (\ref{mar}), (\ref{sum3}), (\ref{kappa}) and 
(\ref{aux}) together we obtain the cluster representation of the
partition function $\Z_T$:
\begin{proposition}
For every $T > 0$ 
\begin{equation}
\Z_T  = 1 + \sum_{n \geq 1} \sum_{\{\Ga_1,...,\Ga_n\}} \prod_{l=1}^n 
K_{\Ga_l}.
\label{clexp}
\end{equation}
Here the summation is performed over all sets of clusters 
$\{\Ga_1,...,\Ga_n\} \neq \emptyset$ for which $\Ga^*_i \cap \Ga^*_j = 
\emptyset$ whenever $i \neq j$. 
\label{p0}
\end{proposition}
As soon as the cluster representation of $\Z_T$ is established, the 
existence of the weak limit measure $\mu = \lim_{T \rightarrow \infty}
\mu_T$ follows by the cluster estimates below and the general arguments 
presented in e.g. \cite{MM}, Chapter 3. 

We conclude the presentation of the expansion by briefly explaining the 
two crucial cluster estimates. The first one is given by 
\begin{proposition}
For every cluster $\Ga$ we have the bound
\begin{equation}
|K_\Ga| \; \leq \; \prod_{\rh \in \Ga} (c_1
|\lambda|^{1/3})^{|\bar\rh|} \prod_{\ga \in \Ga} 
\prod_{(\tau_i,\tau_j) \in \ga} \frac{c_2 |\lambda|^{1/3}}{(|i-j-1|b)^
\delta + 1} 
\label{clest}
\end{equation}
with $|\bar\rh|$ denoting the number of intervals contained in $\rh$, 
constants $c_1, c_2 > 0$ and exponent $\delta > 1$. 
\label{p1}
\end{proposition}
In estimate (\ref{clest}) the factor accounting for the contribution
of chains comes from the uniform upper bound $Ce^{-\Lambda b}$ on 
$|g_b(x|x') - 1|$ (see second factor in (\ref{kappa})), where
$\Lambda$ is the spectral gap of the Schr\"odinger operator of the 
underlying $P(\phi)_1$-process, and $C > 0$. This bound, in its turn, 
is a consequence of the intrinsic ultracontractivity of $e^{-tH}$, 
compare Section \ref{Sect2.1}. The factor accounting for the
contribution of contours comes from an estimate using a generalized 
variant of the H\"older inequality applied to the products over 
$e^{-\lambda W_{\tau_i\tau_j}} - 1$ (see first factor in
(\ref{kappa})). $b$ is finally chosen in such a combination with 
$\lambda$ and $\Lambda$ that the expression (\ref{clest}) results. 

The second fundamental estimate ensuring the convergence of the
cluster expansion is
\begin{proposition}
There is a constant $c > 0$, independent of $\lambda$, and a number 
$0 < \eta(\lambda) < 1$ with $\eta \to 0$ as $\lambda \to 0$, such that
\begin{equation}
\sum_{\Ga: \Ga^* \ni 0 \atop |\bar\Ga| = n} |K_\Ga| \;\leq\;  c \; \eta^n.
\label{conv}
\end{equation}
with $|\bar\Gamma|$ denoting the number of intervals contained in
$\Gamma$ through some contour or chain.
\label{p2}
\end{proposition}
This estimate follows through a procedure of translating the summation 
in the left hand side of (\ref{conv}) into a combinatorial problem and 
resumming over (and counting of) first graphs and then trees. The
contours are assigned vertices and they are linked into graphs
according to the rules connecting them up into clusters. 

\vspace{0.2cm}
So far we assumed free boundary conditions. By an extension of the
argument sketched above also other boundary conditions can be taken
into account, picked from $\X^*$, the subset provided by Theorem
\ref{4.3}. Then an important question is how the limiting measure
depends on the boundary conditions. Uniqueness (in DLR-sense) means 
that for any increasing sequence of real numbers $\{T_n\}$ and any 
corresponding sequence $\{Y_n\} \subset \X^*_{[-T_n,T_n]^c}$, 
$\lim_{n\to\infty} \mathbb{E}_{\mu_{T_n}}[F_B | Y_n] = \mathbb{E}_
\mu[F_B]$, for every bounded $B \subset \R$, and each bounded 
measurable function $F_B$. 
\begin{theorem}
Suppose $V$ is of class (V2) and $W$ satisfies (W2). Then we have the 
following cases:
\begin{enumerate}
\item
If $\alpha > 2$, then whenever the Gibbs measure $\mu$ exists, it is 
unique in DLR sense.
\item
If $\alpha > 1$, then for sufficiently small $|\lambda|$ the limiting 
Gibbs measure $\mu$ is unique in DLR sense whenever the reference
measure is unique.
\end{enumerate}
\label{gibuni}
\end{theorem}
If $\alpha > 2$, $|W^Y_T(X)|$ (given by (\ref{int})) is uniformly 
bounded in $T$, and in paths $X$ and $Y$. This implies that only one 
Gibbs measure can exist, and the argument requires no restriction on 
the values of $\lambda$. For $1 < \alpha \leq 2$ this uniform 
boundedness does not hold any longer and we once again take recourse 
to cluster expansion. 

\vspace{0.2cm}
Having a Gibbs measure at hand, an important aspect in its 
understanding is to see what a typical configuration looks like under 
it. This is answered by 
\begin{theorem}
Under the same conditions as in the previous theorem, with 
$\mu$-probability 1 we have
\begin{equation}
|X_t| \leq C \left(\log (|t| + 1)\right)^{1/(s+1)} + Q(\{X\})
\end{equation}
with a suitable number $C > 0$ and a functional $Q$, independent of
$t$.
\label{tpb}
\end{theorem}
The strategy of proving Theorem \ref{tpb} goes by boosting the typical
behaviour of the reference process explained above to the level of the 
Gibbs measure. First it is shown that for any $a > 0$
\begin{equation}
P\left(\{X \in \X: \max_{0 \leq t \leq 1} |X_t| \geq a\}\right) \; 
\leq C \; e^{-\theta a^{s+1}}
\end{equation}
with appropriate $C, \theta >0$. This can be proven by using
Varadhan's Lemma taken together with the upper bound 
$\exp(-\theta |x|^{s+1})$ for $\psi_0$ (the ground state of the 
Schr\"odinger operator generating the reference process). Then 
Theorem \ref{tpb} comes about by proving that also $C' > 0$ and 
$\theta' > 0$ can be found such that
\begin{equation}
\mu \left(\{X \in \X: \max_{0 \leq t \leq 1} |X_t| \geq a\}\right) \; 
\leq C'  e^{-\theta' a^{s+1}}.
\label{pmu}
\end{equation}
The proof requires once again the use of cluster expansion. 

\vspace{0.2cm}
Finally, we list some additional properties of Gibbs measures for 
(W2)-type pair potentials, useful in various contexts. This case in 
particular covers Nelson's scalar field model, see Section \ref{Sect4}
below.
\begin{theorem}
Let $\mu$ be a Gibbs measure for $W$ satisfying (W2). Suppose $V$ is 
of (V2)-type and $|\lambda|$ is small enough. Then the following hold:
\begin{enumerate}
\item
{\rm{[Invariance properties]}} $\mu$ is invariant with respect to time 
shift and time reflection:
\begin{eqnarray*}
&& \mu \circ \tau_t = \mu, \;\; \forall t \in \R, \;\; \mbox{where} 
\;\;  (\tau_s X)_t = X_{s+t}, \\
&& \mu \circ \vartheta = \mu, \;\; \mbox{where} \;\;  (\vartheta X)_t
= X_{-t}. 
\end{eqnarray*}
\item
{\rm{[Single time distributions]}} The distributions $\vp_T$ under 
$\mu_T$ of positions $x$ at time $t = 0$ are equivalent to $\nu$,
i.e. there exist $C_1,C_2 \in \R$, independent of $T$ and $x$ such 
that
\begin{equation}
C_1 \; \leq \; \frac{d\vp_T}{d\nu}(x) \; \leq \; C_2
\label{aco}
\end{equation}
for every $x \in \R^d$ and $T > 0$. Moreover $\lim_{T \to \infty} 
(d\vp_T/d\nu)(x) = (d\vp/d\nu)(x)$ exists pointwise. 
\item
{\rm{[Single time conditional distributions]}} The conditional 
distributions $\mu_T(\;\cdot\;|X_0 = x)$ converge locally weakly to 
$\mu(\;\cdot\;|X_0 = x)$, for all $x \in \R^d$. 
\item
{\rm{[Mixing properties]}} For any bounded functions $F,G$ on $\R^d$ 
we have on the covariance the estimate
\begin{equation}
|\cov_\mu \; (F_s;G_t)| \;\leq \; {\rm{const}} \; \frac{\sup{|F_s|} 
\sup{|G_t|}}{1 + |t-s|^\beta}
\end{equation}
where $\beta > 0$, $F_s := F(X_s)$, $G_t := G(X_t)$, and the constant 
prefactor is independent of $s,t$ and $F, G$.  
\end{enumerate}
\end{theorem}

\subsection{Existence for pair potential of arbitrary strength}
\label{Sect2.3}

The main restriction in the previous section was that the pair
potential $W$ had to carry a small prefactor $\lambda$. This
restriction is inherent in the cluster expansion. An alternative 
route to the existence of Gibbs measures are compactness arguments; 
the main tool is the concept of uniform domination \cite{Georgii}, 
which in our context reads as follows: 

\begin{definition}
Let $P$, $(\muT)_{T \geq 0}$ be probability measures on $C(\R,
\R^d)$. We say that the family $(\muT)_{T \geq 0}$ is locally 
uniformly dominated by $P$ if the following holds true: For each 
$\varepsilon > 0$ and $S > 0$ there exists $\delta > 0$ such that 
$P(A) < \delta$ implies $ \limsup_{T \to \infty} \muT(A) < 
\varepsilon$ uniformly in sets $A$ depending on $X_s, |s| < S$, 
$(X_s)_{s \in \R} \in C(\R,\R^d)$. 
\end{definition}

The important fact is  that each family $(\muT)_{T \geq 0}$ of 
probability measures that is locally uniformly dominated by a 
probability measure $P$ has at least one cluster point as $T \to 
\infty$ in the topology of local convergence. In order to apply 
this to Gibbs measures we adopt the general set-up from the previous 
section. As a first assumption on the potentials we need

\begin{trivlist}
\item[\bf (A1)] $V$ is Kato-class, i.e. satisfies (V1) from Section 
\ref{Sect2.1}. Moreover, the Schr\"odinger operator $H$ corresponding 
to $V$ has a unique, square-integrable ground state $\psi_0$. 
\item[\bf(A2)] $W$ is extensive, i.e. there exists $C_{\infty} > 0$ 
such that
\begin{equation} \label{Cinfty}
\int_{-\infty}^{\infty} \sup_{x,y \in \mathcal \R^d} |W(x,y,|s|)| 
\, ds < C_{\infty}. 
\end{equation}
\end{trivlist}
Comparing with the previous section, we find that (W2) implies 
$(A2)$. 

Let $P$ be the measure of the stationary $P(\phi)_1$-process as given 
in Section \ref{Sect2.1}, and let $W_{[-T,T]}$ be given by
(\ref{inn}). We will use finite time interval Gibbs measures with free 
boundary conditions as approximants for our infinite time interval 
Gibbs measures, i.e. we put 
$$ d \muT = \frac{1}{\Z_T} e^{-W_{[-T,T]}(X)} \, dP.$$
Using the concept of local uniform domination, it is now possible to 
prove 

\begin{proposition} \cite{betz thesis}
Assume (A1) and (A2). Suppose that for each $\varepsilon > 0$ there 
exists $R>0$ such that 
\begin{equation} \label{U}
\qquad \muT(|X_0| > R) < \varepsilon
\end{equation}
uniformly in $T > 0$. Then there exists an (infinite time interval)
Gibbs measure for the potentials $V$ and $W$ and the reference measure 
$\mathcal{W}$ (Wiener measure).
\end{proposition}

We have thus reduced the problem to proving (\ref{U}). For this we
need some further assumptions.

\begin{trivlist}
\item[\bf (A1')] 
In addition to (A1) suppose $\psi_0 \in L^1(\R^d)$.
\end{trivlist}
Condition (A1') is not very restrictive;  in many cases $\psi_0$ 
decays exponentially at infinity. The additional condition on $W$ will 
be more restrictive and requires some preparations to formulate. Let 
$C^{(0)}(\R,\R^d)$ denote the space of functions which are continuous 
with the possible exception of the origin but have left and right hand 
side limits there. For $\tau > 0$ we define the map  
\begin{equation} \label{theta}
\theta_{\tau}: C(\R, \R^d) \to C^{(0)}(\R,\R^d), \quad 
(\theta_{\tau} X)_t = \left\{ 
\begin{array}{ll} 
    X_{t+\tau} & \mbox{if } t \geq 0,\\
    X_{t-\tau} & \mbox{if } t < 0.
\end{array} \right.
\end{equation}
With $E_0 = \inf\mathrm{Spec}(H)$ as before, and $H$ the Schr\"odinger 
operator corresponding to the $P(\phi)_1$-process $P$, put 
\begin{equation} \label{alpha}
    \alpha = \liminf_{|x| \to \infty} V(x) - E_{0} \leq \infty.
\end{equation}
Our assumption on $W$ now reads
\begin{trivlist}
\item[\bf (A2')] In addition to (A2), we assume that there exist $D \geq
  0$ 
and $0 \leq C < \alpha$ such that
\begin{equation} \label{cond}
    -W_{[-T,T]} (X) \leq -W_{[-T,T]}(\theta_{\tau} X) + C \tau + D
\end{equation} 
for all $T, \tau >0$ and all $X \in C(\R,\R^d)$.
\end{trivlist}
In words, (A2') means that we can control, uniformly in $T$,  the 
change of energy induced by cutting out a piece of the path $X$ around 
$t=0$ and gluing the remaining pieces  together again. If we have
finite interaction energy between the positive and the negative
half-line, i.e.
\begin{equation} \label{leftright}
\sup_{X \in C(\R, \R^d)} \left| \int_{-\infty}^0 ds 
    \int_0^{\infty} dt \,  W(X_t,X_s,|t-s|) \right| < \infty,
\end{equation}
then (A2') holds with $C=0$. In particular, (\ref{leftright}) holds 
when $W$ fulfills (W2) with $\alpha > 2$. (\ref{leftright}) is,
however, not necessary for (A2'), and part of the interest in 
condition (A2') is that it also covers cases where (\ref{leftright}) 
is not met. Some sufficient conditions for (A2') are given in 
\cite{Betz}.
 
\begin{theorem} \label{compactness existence}
Assume (A1') and (A2'). Then (\ref{U}) holds, and consequently an 
infinite volume Gibbs measure $\mu$ for the potentials $V$ and $W$, 
and reference measure $\mathcal{W}$ exists.
\end{theorem}
The theorem above does not make any statement about uniqueness. 
However, in conjunction with (2) of Theorem \ref{gibuni} it leads to
\begin{corollary}
Provided (W2) with $\alpha > 2$ holds, and $V$ satisfies (A1') and
(A2'), a unique Gibbs measure exists supported by $\X$. 
\end{corollary}
Hariya \cite{H} arrives at a similar result under different
hypotheses.

The proof of (\ref{U}) relies on the equality  
\begin{equation} \label{3.2.1}
\muT(|X_0| > R) = \frac{1}{\Z_T} \int_{|y| > R}  \psi_0^2(y) \EE_{P} 
\Big[e^{-W_{[-T,T]}} \Big| X_0=y \Big] dy.
\end{equation}
We first prove 
\begin{equation} \label{1/phi}
\frac{1}{\Z_T} \EE_{P} \left[ e^{-W_{[-T,T]}} \Big| X_0 = y \right] 
\leq \frac{\mathrm{const}}{\psi_0(y)}
\end{equation}
and then use (A1') in order to obtain  (\ref{U}). To get an idea about 
the proof of the latter inequality, note that  (\ref{1/phi}) involves 
expectation with respect to a Markov process conditioned at its 
`midpoint' $t=0$. For making use of the strong Markov property of $P$, 
we flip the negative time axis to the right and obtain a Markov 
process with a doubled state space $\R^{2d}$, now conditioned on its 
starting point. Now we start the new process in $y \in \R^{2d}$ and 
stop it when it reaches the ball $B_r$ around zero with radius $r$. By 
the properties of the $P(\phi)_1$-process, the stopping time $\tau_r$ 
the process needs to reach $B_r$ is exponentially integrable. More 
explicitly, $\EE^x_P[\exp(\beta \tau_r)] < \infty$ if $\beta <
\alpha$, and the expectation value grows with the starting point $x$ 
like $1/\psi_0(x)$ as $x \to \infty$. Condition (A2') is now
tailor-made to ensure that the energy $\tilde W_{[-T,T]}$ acquired by 
a (flipped) path $X$ on its way down to the $B_r$ is no larger than 
$\exp(C \tau_r + D)$. Together with the strong Markov property and 
some technical estimates, this yields  (\ref{U}).

\subsection{Phase transition}
\label{Sect2.4}

In one-dimensional statistical mechanical sytems the entropy
increases as $\log T$. To have a phase transition the interaction
energy for the paths $\{X_t,\; -T \leq t \leq 0\}$ and $\{X_t, \;
0 \leq t \leq T\}$ must be at least comparable. Transcribed to the
Gibbs measures under study this means
\begin{equation}\label{2.1}
W (x,x',t) \cong |t|^{-\gamma} \quad \textrm{for large}\; |t|
\end{equation}
with $1<\gamma\leq2$. The lower bound on $\gamma$ is needed for 
having the energy extensive. To carry out a proof more specific 
assumptions will be needed. We set $d=1$. For the external 
potential we choose a double well potential of the form
\begin{equation}\label{2.2}
V(x) = \beta (x^4-x^2)\,, \quad \beta>0\,.
\end{equation}
In fact, as long as $V(x) = V(-x)$, a general class of double well
type potentials can be handled. The pair interaction is quadratic,
\begin{equation}\label{2.3}
W(x,x',t)= \alpha \rho(t) \frac{1}{2}(x-x')^2\,, \alpha >0\,,\quad
\rho(t) = (1+|t|)^{-\gamma}\,.
\end{equation}
Since we rely on comparison inequalities, the interaction needs to be 
quadratic, at least at the present stage of understanding. Thus the 
only non-Gaussian piece of the Gibbs measure is 
$\exp[-\beta \int^T_{-T} (X_t)^4 dt]$. Let
$\langle\;\cdot\;\rangle_{b,T}$ be the expectation of the Gibbs
measure for the potentials $V$ and $W$ from (\ref{2.2}),
(\ref{2.3}), with the pinned boundary conditions $X_{-T}=b=X_T$,
$b \in \mathbb{R}$. Then, for $b>0$, $\langle X_0 \rangle_{b,T}
\geq 0$ and $\langle X_0 \rangle_{b,T}$ is decreasing in $T$.
Hence the limit
\begin{equation}\label{2.4}
\lim_{T \to \infty} \langle X_0 \rangle_{b,T} = \langle X_0
\rangle_{b,\infty}
\end{equation}
exists. 
\begin{theorem} Let $V, W$ be as in (\ref{2.2}), (\ref{2.3})
and fix $1< \gamma \leq 2$.\label{t1} If $b>0$, then there exist
$\alpha, \beta, m^\ast > 0$ such that
\begin{equation}\label{2.5}
\langle X_0 \rangle_{b,\infty} \geq m^\ast.
\end{equation}
\end{theorem}
By symmetry, $\langle X_0 \rangle_{-b,\infty} = - \langle X_0
\rangle_{b,\infty}$. Thus there must be at least two distinct
extreme Gibbs measures for the same interaction. Most likely there 
are no others, but this problem has not been approached yet.

The strategy of proof is to reduce the bound in (\ref{2.5}) to a
corresponding one for a one-dimensional Ising spin system with
long-range ferromagnetic pair interaction, for which the famous
proofs of Dyson \cite{dys} and of Fr\"{o}hlich and Spencer
\cite{fs} on the existence of long range order are available. The
reduction is based on ferromagnetic type inequalities. With the
block variables
\begin{equation}\label{2.6}
\phi_j = \frac{1}{\delta}
\int_{(j-\frac{1}{2})\delta}^{(j+\frac{1}{2})\delta} X_t
dt\,,\quad j \in \mathbb{Z}\,,
\end{equation}
by Griffiths II we obtain that $\langle X_0 \rangle_{b,\infty}
\geq c_{\tiny\textrm{G}} \langle\phi_0 \rangle^c_{b,\infty}$, where
$\langle\; \rangle^c$ is a Gibbs measure over $\mathbb{Z}$ with
long range interaction for the continuous spin variables $\phi_j$,
and $c_{\tiny\textrm{G}}>0$. Secondly, the Wells inequality, see
\cite{Sim2} in the case of stochastic processes, \cite{bri} implies 
that $\langle\phi_0 \rangle^c_{b,\infty} \geq c_{\tiny\textrm{W}} 
\langle\sigma_0 \rangle_{+,\infty}$ with $c_{\tiny\textrm{W}} >0$. 
Here $\langle \; \rangle_{+,\infty}$ is an Ising spin system,
$\sigma_0 = \pm$, with ferromagnetic interaction which decays as 
$|i-j|^{\gamma}$ for large $|i-j|$ and $+$ boundary conditions. The 
complete proof is given in \cite{os}, where also explicit bounds for 
the phase diagram are discussed.

\section{A central limit theorem}
\label{Sect3}

In this section we study the case where $V=0$, i.e. we consider
\begin{equation} \label{muT}
\muT = \frac{1}{\Z_T} \exp \left( - \int^T_{-T} \int^T_{-T}
W(X_t-X_s,t-s) \, dt \, ds \right) \, \W^0_T.
\end{equation}
Here, $\W^0_T$ is two-sided Brownian motion in $[-T,T]$ pinned at $0$ 
at $t=0$. The interaction depends only on the increments
$X_t-X_s$. Provided $W$ has a decent decay in the $t$-variable, one 
would thus expect a functional central limit theorem to hold, i.e. 
after rescaling the path measure $\muT$ should look like Brownian 
motion with some effective diffusion matrix $D$. Such a general result 
is not available. In case $t \mapsto W(\cdot,t)$ decays exponentially, 
one can use Dobrushin's theory of one-dimesional spin systems 
\cite{Do73, Do73a} to establish exponential mixing of the increment 
process \cite{Spohn}. This implies the central limit theorem for 
$X_t$ properly rescaled. Our approach is less restrictive in terms 
of decay conditions, but assumes $W$ to be of the special form
\begin{equation}\label{def W}
W(x,t)= -\frac{1}{2} \int|\widehat\rho (k)|^2 e^{ik\cdot x}
e^{-\omega(k)|t|} \frac{1}{2\omega(k)} dk
\end{equation}
with 
\begin{equation}
\omega(k)\geq 0, \quad \omega(k)=\omega(-k), \quad 
\mbox{ and }\quad \widehat\rho(k)=\widehat\rho(-k)^\ast. 
\label{cond1}
\end{equation}
In addition, we assume 
\begin{equation}
\int|\widehat\rho(k)|^2 (\omega^{-1}+\omega^{-2}+\omega^{-3})dk <
\infty. 
\label{cond2}
\end{equation}
(\ref{cond2}) is in fact a (mild) decay condition. For example, 
if $d=3$, $\om(k) = |k|$ and $\widehat\rho$ is compactly supported,
then the most stringent condition is $\int |\widehat\rho|^2
\omega^{-3} d^3 k < \infty$, which corresponds to a decay of $W$ as 
$$|W(x,t)|\leq c (1+|t|^{3+\delta})^{-1},$$ 
for some $\delta>0$. The above choice of parameters represents a 
physically relevant model, see (iv) of Section \ref{Sect4}. 

\begin{theorem}\label{clt}

Define $\muT$ as in (\ref{muT}) with $W$ given by (\ref{def W}). 
\begin{itemize} 
\item[(i):]  $\muT$ converges to a measure $\mu$ as $T \to \infty$ in 
the topology of local convergence. 
\item[(ii):] The stochastic process $X_t$, $t \geq 0$, induced by 
$\mu$ satisfies a functional central limit theorem
$$
\lim_{\varepsilon \to 0} \sqrt{\varepsilon} X_{t/\varepsilon} = 
\sqrt{D}B_t
$$
in distribution, where $0 \leq D \leq 1$ as a $d \times d$ matrix, 
and $B_t$ is standard Brownian motion.
\item[(iii):] In addition to (\ref{cond1}),(\ref{cond2}) suppose
\begin{equation} \label{cond3}
\int |\widehat{\rho}(k)|^2 |k|^2 \left( \om^{-2} + \om^{-4}\right) 
\, dk < \infty.
\end{equation}
Then $D > 0$. 
\end{itemize}
\end{theorem}

In the remainder of this section, we will give an outline of the proof 
of Theorem \ref{clt}. A full account is \cite{BS04}. We will do the 
proof in three steps.
\begin{itemize}
\item[(1) ] We use the special form (\ref{def W}) of $W$ in order to 
linearize the interaction in (\ref{muT}) by introducing an auxiliary 
Gaussian process. As a result, we will prove (i) above, and the
stochastic process $X_t$ under $\mu$ is driven by a reversible Markov 
process $\eta_t$. 
\item[(2) ] In the so obtained representation, we use the by now 
well-established technique of Kipnis and Varadhan \cite{KV}; we write 
$X_t$ as the sum of a martingale and an additive functional of
$\eta_t$. $X_t$ is then the sum of two 
martingales and a negligible process, and the martingale central limit 
theorem applies, proving (ii).
\item[(3) ] In order to show that the diffusion is nondegenrate, we 
rely on an idea of Brascamp, Lebowitz and Lieb \cite{Brascamp}, which 
in the present context has been employed before \cite{Spohn}.
\end{itemize}
To carry out step (1), let $\cK_0$ be the real Hilbert space obtained
by completing the subspace of $L^2(\R^d)$ on which the inner product
given by 
\begin{equation} \label{cK}
\skalprod[\cK_0]{a}{b} = \int \widehat{a}(k) 
\frac{1}{2\om(k)}\widehat{b}(k)^{\ast} \, dk
\end{equation}
is finite. Let $\G$ be the path measure of the infinite dimensional
Ornstein-Uhlenbeck process with mean $0$ and covariance 
$$ \EE_\G[ \phi_{s}(a) \phi_{t}(b)] =\int \widehat{a}(k)
    \frac{1}{2\om(k)}e^{-|t-s|\om(k)} \widehat{b}(k)^\ast \, dk
    \qquad (a,b \in \cK_0).$$
There exists a Hilbert space $\cK \supset \cK_0$ such that $\G$ is a 
reversible Gaussian Markov process with values in $\cK$ and 
continuous paths. The reversible measure $\g$ is the Gaussian measure 
on $\cK \ni \phi$ with mean zero and covariance 
$$\EE_{\g}[\phi(a) \phi(b)] =  \skalprod[\cK_0]{a}{b}.$$ 
For $x \in \R^d$, let $\tau_x$ be the shift by $x$ on $\cK$, i.e. 
$(\tau_x\phi)(a) = \phi(\tau_x a)$ and $\tau_x a(y) = a(y-x)$. 
More generally, for $f \in L^2(\g)$, $(\tau_x f)(\phi) =
f(\tau_x\phi)$

For $T>0$ we put 
\begin{equation} \label{P-measure}
\P_T = \frac{1}{\Z_T} \exp \left(- \int_{-T}^T \tau_{X_s}\phi_s
( \rho) \, ds\right) \, \W^0 \otimes \G.
\end{equation}
With $\P_T$ we achieved our first goal, the linearization of the 
interaction: Indeed, for functions $F$ depending on $x$ only, 
$$ \EE_{\P_T}[F] = \EE_{\muT}[F],$$
as can be seen by carrying out the Gaussian integration. $\P_T$ is the 
measure of a Markov process, more specifically a $P(\phi)_1$-process 
with state space $\R^d \times \cK$. The role of the Schr\"odinger 
operator is now played by  
\begin{equation} \label{generator}
Hf(x,\phi) = - \frac{1}{2} \Delta f(x,\phi) + H_{\mr f}f(x,\phi) + 
V_{\rho}(x,\phi)f(x,\phi),
\end{equation}
where $H_{\mr f}$ is the generator of $\G$ and $V_{\rho}(x,\phi) 
= \tau_x\phi(\rho)$. The semigroup $\Pi_T$ generated by $H$ is 
strongly continuous on $C_0(\R^d, L^2(\g))$. More importantly, it is 
also strongly continuous on the Hilbert space $\cT$ of functions 
that are invariant under shift over the $x$-variable. Explicitly, 
$\cT$ is the image of $L^2(\g)$ under the operator 
$$
U:L^2(\g) \to C(\R,L^2(\g)), \quad Uf(x,\phi) = \tau_xf(\phi),
$$
equipped with the scalar product
\begin{equation} \label{scalar product in T}
\skalprod[\cT]{f}{g} = \EE_{\g}[(U^{-1}f) (U^{-1}g)^\ast]= 
\skalprod[L^2(\g)]{U^{-1}f}{U^{-1}g}. 
\end{equation}
$H$ is self-adjoint on $\cT$, and (\ref{cond2}) implies
$$\norm[\cT]{\Pi_T1}^2 \leq  C \skalprod[\cT]{1}{\Pi_T1}^2.$$
Now from spectral theory we obtain

\begin{theorem} \label{existence of ground state}
The infimum $E_0$ of the spectrum of $H$ acting in $\cT$ is an 
eigenvalue of multiplicity one. The corresponding eigenfunction 
$\Psi \in \cT$ can be chosen strictly positive.
\end{theorem}
An alternative proof of Theorem \ref{existence of ground state}, 
using a completely different method, can be found in \cite{Froehlich}.

It is now easy to identify the infinite volume limit of the families 
$\P_{T}$ and $\N_{T}$. Let  $\P$ be the probability measure on paths 
$(X_t,\phi_t)_{t \in \R}$ determined by 
\begin{equation} \label{P}
\EE_{\P}(f) = e^{2TE_0} \EE_{\W^0 \otimes \G} 
\left[ \Psi(X_{-T},\phi_{-T}) e^{- \int_{-T}^T \tau_{X_s}
\phi_s( \rho) \, ds} \Psi(X_{T},\phi_{T}) f \right]
\end{equation}
for functions $f$ depending only on $X_t,\phi_t$ with $|t|<T$. Above, 
$\W^0$ is the measure of two-sided Brownian motion or, equivalently, 
Wiener measure conditioned on $X_0 = 0$. Let $\mu$ be the measure $\P$ 
when applied to functions of $x$ only. Then $\P$ is the measure of a 
Markov process with generator $L$ acting as
\begin{equation} \label{process generator}
Lf = - \frac{1}{\Psi}(H-E_{0})(\Psi f).
\end{equation}
$\P_T \to \P$ in the topology of local convergence, and by integrating 
out the Gaussian field, $\mu_T \to \mu$. The $\cK$-valued  process 
$ \eta_t = \tau_{X_t} \phi_t$ is reversible with reversible measure 
$(U^{-1}\Psi)^2 \g$, and its generator is unitarily equivalent to $L$. 

Let  $\gamma \in \R^d$ be fixed, and $h_\gamma(x) = \gamma \cdot x$. 
Then $L(h_\gamma) = j(\eta)$ with 
\begin{equation}\label{Lf=j}
j = U^{-1}(\gamma \cdot \nabla_x \ln \Psi) \in L^2(\g).
\end{equation}
Since the result of the generator $L$ of process $\P$ applied to 
$\gamma \cdot x$ is a function of $\eta$, only $\eta_t$ influences the 
behavior of $\gamma \cdot X_t$, i.e. $X_t$ is driven by $\eta_t$. Step 
one is completed.

Next we write
\begin{equation} \label{sum of two martingales}
\gamma \cdot X_t =  M_t +  \int_{0}^{t} Lh_\gamma(X_s, \phi_s)) \, ds
\end{equation}
with
$$ M_t = \gamma \cdot X_{t} - \int_{0}^{t} Lh_\gamma(X_s, \phi_s) \, ds  =
\gamma \cdot X_t - \int_0^t j(\eta_s) \, ds.$$
Then $M_t$ is a martingale with stationary increments and quadratic 
variation $|\gamma|^2t$, and 
$$ \int_{0}^{t} Lh_\gamma(X_s, \phi_s)) \, ds =  \int_0^t j(\eta_s) \, ds $$
is an additive functional of $\eta_t$ satisfying the assumptions of
\cite{KV}. It is thus the sum of a martingale $N_t$ with stationary 
increments and a negligible process. Now the martingale central limit
theorem proves Theorem \ref{clt} (ii) and finishes step 2.

In principle, it could happen that $M_t$ and $N_t$ are strongly
dependent and cancel each other. Then the diffusion matrix $D$ would
be zero and $X_t$ would behave subdiffusively. We already know the
central limit theorem holds with diffusion matrix $D \geq 0$. Thus it
is enough to investigate 
\begin{equation} \label{DD}
\lim_{t \to \infty} \frac{1}{t} \EE_\mu[(\gamma \cdot X_t)^2] = 
\skalprod[\R^d]{\gamma}{D \gamma}.
\end{equation}
It turns out that 
\begin{equation} \label{diffusion constant}
\skalprod[\R^d]{\gamma}{D \gamma} = 
|\gamma|^2 - 2 \skalprod[\cT]{\gamma \cdot \nabla_x\Psi}{(H-E_0)^{-1}
\gamma \cdot \nabla_x\Psi}.
\end{equation}
The standard technique is to turn (\ref{diffusion constant}) into a 
variational problem and find a reasonably explicit lower bound to the 
variational functional. We did not succeed in carrying out the second 
step of this procedure. Instead, we show directly that
\begin{equation} \label{est}
\EE_{\mu}[(\gamma \cdot X_t)^2] \geq c |\gamma|^2 |t|
\end{equation}
for some $c > 0$, by using ideas from Brascamp {\it et al} 
\cite{Brascamp} originally developed to study fluctuations for 
anharmonic lattices. Together with (\ref{DD}) this immediately shows 
$D\geq c$.

\section{Applications and open problems} 
\label{Sect4}

The scheme outlined so far is a probabilistically natural way of
constructing through the limit $T \to \infty$ stationary stochastic 
processes with continuous sample paths. Moreover, specific choices of 
$V$ and $W$ correspond to particular applications on which there is 
already a large body of literature using a variety of methods. Very 
roughly, and as far as we are aware of, the applications originate 
from three distinct corners of low energy physics.

\vspace{0.2cm}
\noindent
\textit{i) Self-avoiding random walks.} Polymers with interaction
due to excluded volume is an important statistical mechanics
topic, in particular because of the connections with critical
phenomena \cite{ffs}. It is tempting to
model the free polymer as Brownian motion and the excluded volume
through an interaction of the form (\ref{1.7}). Note, however, that by 
the nature of the interaction there is no decay in $t$. In particular
the energy is not extensive. Thus, while the energy depends only
on the increments, for large $T$ the statistical properties of the
self-avoiding polymer are qualitatively different from a free
random walk. One conjecture is that the self-similar scaling
theory is obtained from the ultraviolet limit. So far most of
the mathematical effort went into constructing the limit measure
\cite{west,bolt}. But it is not obvious how to
extract scale invariant properties from this measure. In fact,
self-similarity is now established through lace expansion and other 
methods \cite{ms}. The link between the two approaches remains 
unexplored.

\vspace{0.2cm}
\noindent
\textit{ii) Statistical hydrodynamics.} There is general agreement
that fully developed turbulence should be described by a suitable
measure over divergence free vorticity fields. One attempt to write
down such a measure is to assume that the velocity field
$\omega(x)= \nabla \wedge u(x)$ is concentrated along Brownian
curves $X_t \in \mathbb{R}^3$ \cite{chor}. Under the
Eulerian incompressible flow, the kinetic energy $\frac{1}{2} \int
u(x)^2 d^3 x$ is conserved. Thus it seems natural to use it as
energy in the Gibbs measure. This yields the formal expression
\begin{equation}\label{3.11}
\mathcal{E}(X)= \int^T_{-T} \int^T_{-T} \frac{1}{|X_t - X_s|} dX_t
\cdot dX_s\,.
\end{equation}
In order to have $\exp[-{\cal E}(X)]$ as a well-defined random
variable, \cite{fla} required the condition that the Coulomb potential 
in (\ref{3.11}) is smoothened such that it has a finite electrostatic
energy. 

\vspace{0.2cm}
Our own investigations mostly draw on applications in quantum
mechanics. Since upon Wick rotation the free Schr\"{o}dinger
equation turns into the diffusion equation, Brownian motion as 
{\it a priori} measure is in fact forced by the problem. Several
interesting cases can be distinguished.

\vspace{0.2cm}
\noindent
\textit{iii) Electron coupled to the quantized radiation field.}
Upon Wick rotation the free Maxwell field is isomorphic to a
stationary infinite-dimensional Ornstein-Uhlenbeck process, see
Section \ref{Sect3}, for the transverse vector potential $A(x,t)$. It 
has the covariance
\begin{eqnarray}\label{3.22}
\mathbb{E}\big[A_{\alpha}(x,t)A_{\beta}(x',t')\big] & = &\int d^3k
\frac{1}{2\omega(k)} e^{-\omega(k)|t-s|} e^{ik
\cdot(x-x')}(\delta_{\alpha\beta}-|k|^{-2}k_\alpha k_\beta)
\nonumber\\
& = & W_{\alpha\beta}(x-x',t-s)
\end{eqnarray}
$\alpha, \beta=1,2,3$. The dispersion relation of the Maxwell
field is
\begin{equation}\label{3.3}
\omega(k)=|k|\,.
\end{equation}
Within the Euclidean framework, the electron is governed by the
Hamiltonian
\begin{equation}\label{3.4}
H= \frac{1}{2}\big(-i \nabla_x -eA(x,t)\big)^2\,,
\end{equation}
on ignoring the electron spin. The units are such that $\hbar=1$,
$c=1$, mass of electron $m=1$; $e$ is the charge of the electron
expressing the strength of coupling to the Maxwell field. We use
the Feynman-Kac-Ito formula for the propagator for $H$
\cite{Sim2}. Then the joint $X_t$ and $A(x,t)$ path measure 
is given by 
\begin{equation}\label{3.4a}
\exp\left(-i e\int_{-T}^T A (X_t,t) \cdot dX_t \right) \; 
\mathcal{W}(X) \otimes \mathcal{G}, 
\end{equation}
where $\mathcal{G}$ is the Gaussian measure of the $A$-field with
covariance (\ref{3.22}). Note that $\nabla \cdot A(x,t)=0$ almost
surely. Since the exponent is linear in $A$, the averaging over the
Ornstein-Uhlenbeck process can be done explicitly. This results in
a finite volume Gibbs measure with energy
\begin{equation}\label{3.5}
\mathcal{E}(X)= \frac{1}{2} \int^T_{-T} \int^T_{-T} dX_t \cdot
W(X_t - X_s, t-s) dX_s\,.
\end{equation}
This is of the form (\ref{1.6}) and should be read as a double 
Ito stochastic integral.

$W$ is singular on the diagonal, roughly $W_{\alpha\beta}(x,t) =
\delta_{\alpha\beta}(x^2 + t^2)^{-1}$. Thus it is necessary to
smear out the charge distribution which leads to the regularized
version
\begin{equation}\label{3.6}
W^\rho_{\alpha\beta}(x,t)= \int d^3k|\widehat{\rho}(k)|^2
\frac{1}{2\omega(k)}e^{-\omega(k)|t|} e^{ik \cdot x}
(\delta_{\alpha\beta}-|k|^{-2}k_\alpha k_\beta).
\end{equation}
Here $\widehat{\rho}$ is rotation invariant, decays rapidly for
large $|k|$, and $\widehat{\rho}(0) = (2 \pi)^{-3/2}$ by charge
normalization. A problem which appears to be very challenging, is to 
establish that, for fixed $T$ and $X_{-T}=0=X_T$, the Gibbs measure 
for the energy (\ref{3.5}) is well defined. In other words, with a 
smoothening as in (\ref{3.6}) we would like to study the sequence of 
Gibbs measures as $\widehat{\rho}(k) \to (2\pi)^{-3/2}$ pointwise 
(ultraviolet or point charge limit). In favorable cases the existence 
of the limit can be shown by suitable centering and by possibly adding
other counter terms. Such  a procedure seems unlikely to work in the 
present context. Thus the ultraviolet limit has to be linked with a 
change of the diffusion coefficient $D$ of the underlying Wiener 
process $\mathcal{W}$ (= mass renormalization). We expect $D \to 
\infty$ in this limit.

\vspace{0.2cm}
\noindent
\textit{iv) Quantum particle coupled to a scalar Bose field.} This
model was studied by Nelson \cite{nel} in the context of energy
renormalization. The Bose field translates to the scalar field
$\phi(x,t)$, which again is an infinite-dimensional
Ornstein-Uhlenbeck process this time with covariance
\begin{eqnarray}\label{3.7}
\mathbb{E}\big[\phi(x,t)\phi(x',t')\big] & = & \int d^3
k|\widehat{\rho}(k)|^2 \frac{1}{2\omega(k)} e^{-\omega(k)|t-t'|} 
e^{ik \cdot(x-x')}\nonumber\\ 
& = & W(x-x',t-t').
\end{eqnarray}
The quantum particle ``sees'' $\phi$ as a fluctuating electrostatic
potential. Thus the Hamiltonian becomes $H = -\frac{1}{2}\Delta + e
\phi(x,t)$. Then through the Feynman-Kac formula the path measure,
jointly for $X_t$ and $\phi(x,t)$, is given by
\begin{equation}\label{3.7a}
\exp \left(-e\int_{-T}^T \phi (X_t,t)dt \right) \; 
\mathcal{W}^0(X) \otimes \mathcal{G},
\end{equation}
which has the structure of a $P(\phi)_1$-process, since the
\textit{a priori} measure is Markovian and the energy is local 
in time. The only difference to our discussion in Section
\ref{Sect2.1} is that $\mathbb{R}^d$ is replaced by the state 
space $\mathbb{R}^d \times \mathcal{K}$, compare with the 
discussion preceding (\ref{generator}).

The exponent in (\ref{3.7a}) is linear in $\phi$. Thus we can
perform the integration over $\phi$ resulting in the following
path measure for $X$,
\begin{equation}\label{3.7b}
\frac{1}{Z(T)} \exp\big[-\int_{-T}^T V(X_t) dt + \frac{e^2}{2}
\int_{-T}^T \int_{-T}^T W(X_t -X_s, t-s)dtds \big] \; 
\mathcal{W}^0,
\end{equation}
where we added an external potential $V$. Thus the Nelson model
naturally yields Gibbs measures of the form studied in Sections 
\ref{Sect2} and \ref{Sect3}. In fact, the Nelson model was our 
source of motivation for studying Gibbs measures over Brownian 
motion. The existence of the infinite volume limit can be deduced 
from (\ref{P}) which requires that 
\begin{equation}\label{3.8}
\int dk |\widehat{\rho}(k)|^2 (\omega(k)^{-3}+\omega(k)^{-1}) <
\infty.
\end{equation}
We can also use the cluster expansion which holds provided $e^2$
is sufficiently small and
\begin{equation}\label{3.8a}
\int dk |\widehat{\rho}(k)|^2
(\omega(k)^{-1}+\omega(k)^{-2-\delta}) < \infty
\end{equation}
for some $\delta>0$. Since it is possible to express the ground state 
of the Nelson model directly in terms of data of these Gibbs measures,  
given the existence of the infinite time interval measures we have
a useful tool at hand for studying qualitative properties of the
ground state. We refer to \cite{BHLMS} for details.

The Nelson model, in the case of massless bosons $\omega(k)=|k|$,
is both ultraviolet and infrared divergent. The ultraviolet
divergence is mild and can be handled by energy renormalization.
This is the content of the famous work \cite{nel}, which uses
exclusively functional analytic methods. Somewhat surprisingly, no
one has succeeded in a proper transcription of Nelson's results into 
the framework of path measures. The infrared divergence translates 
into a somewhat unexpected feature of the joint 
$\big(X_t,\phi(x,t)\big)$ process. From (\ref{3.8a}) and suitable 
conditions on $V$, we infer that the infinite volume Gibbs measure 
exists. However, the limiting procedure changes the situation seen 
by the \textit{a priori} measure dramatically. For instance, the 
$t=0$ joint distribution is not absolutely continuous with respect 
to the $t=0$ projection of the \textit{a priori} distribution. One 
way to cope is to introduce a suitable shifted Gaussian measure 
which takes on the role of a new \textit{a priori} measure making 
the model infrared regular. We refer for more details to
\cite{LMS1,LMS2}.

\vspace{0.2cm}
\noindent
\textit{v) The polaron.} Physically the polaron is an electron
coupled to the optical mode of an ionic crystal. It can be viewed
as a particular case of the Nelson model with the choice
$\omega(k) =\omega_0$ and $\widehat{\rho}(k) =|k|^{-1}$. Then
\begin{equation}\label{3.9}
W(x,t) = - \frac{\alpha}{|x|} e^{-\omega_0|t|}.
\end{equation}
Here $\alpha >0$ and subsumes all dimensional coupling
coefficients. The ground state energy of the polaron is defined
through
\begin{equation}\label{3.10}
E_\textrm{\tiny g}(\alpha) = - \lim_{T \to \infty}\frac{1}{T}\log
Z(T,\alpha).
\end{equation}
For small $\alpha$ one can use perturbation theory in $\alpha$.
For large $\alpha$ Pekar \cite{pek} developed an approximate
strong coupling theory. Thus the challenge was to have reliable
predictions at moderate values of $\alpha$, which turned out to be
difficult. Feynman \cite{f,fh} had the
insight from functional integration and used a quadratic
functional as upper variational bound. Optimizing the quadratic
form yields $E_{\tiny\textrm{g}}(\alpha)$ roughly 2\% away from 
Pekar's result and even better at smaller values when compared with
machine computations. The strong coupling (Pekar) limit of the
ground state energy has been established by Donsker and Varadhan
\cite{dv} using functional integration, and by Lieb and Thomas
\cite{lt} using functional analytic methods.

A long standing open problem is to obtain a corresponding result
for the effective mass $m(\alpha)$. In fact, as shown in 
\cite{Spohn}, $m(\alpha) = D(\alpha)^{-1}$ with $D(\alpha)$ the
diffusion coefficient in Section \ref{Sect3} with the specific choice 
(\ref{3.9}) for $W$. On heuristic grounds one can guess the behavior of 
$D(\alpha)$ for large $\alpha$ and relate it to Pekar's variational 
problem \cite{Spohn}. A proof is missing with the exception of
\cite{Peter} in the simplification where Brownian motion on
$\mathbb{R}^3$ is replaced by Brownian motion on the circle.

\end{document}